\documentclass{elsart}
\usepackage{graphics}
\usepackage{graphicx}
\usepackage{epsfig}
\usepackage{color}

\usepackage{amssymb}
\begin{document}
\begin{frontmatter}
\title{Conversion efficiency and luminosity for
gamma-proton colliders based on the LHC-CLIC or LHC-ILC
QCD Explorer scheme}

\author[ankara,nigde]{H. Aksakal\corauthref{label2}}
\corauth[label2]{Corresponding author. Tel: +903122126720, fax: +903122232395}
\ead{aksakal@science.ankara.edu.tr},
\author[ankara]{A.K. Ciftci},
\author[ankara,nigde]{Z. Nergiz},
\author[cern]{D. Schulte},
\author[cern]{F. Zimmermann}

\address[ankara]{Department of Physics, Faculty of Science, Ankara University, 06100 Tandogan, Ankara, Turkey}
\address [cern]{CERN, 1211, Geneva 23 ,Switzerland }
\address[nigde]{Department of Physics, Faculty of Art and Science, Nigde University, 51200 Nigde, Turkey}

\begin{abstract}
Gamma-proton collisions allow unprecedented investigations of the
low x and high $Q^{2}$ regions in quantum chromodynamics. In
this paper, we investigate the luminosity for
``ILC"$\times$LHC ($\sqrt{s_{ep}}=1.3$ TeV) and ``CLIC"$\times$LHC
($\sqrt{s_{ep}}=1.45$ TeV) based $\gamma p$
colliders. Also we determine the laser properties required for high
conversion efficiency.
\end{abstract}

\begin{keyword} Photon-Proton Collisions , Luminosity
\PACS 13.60.Fz; 41.75.-i; 42.55.-f
\end{keyword}
\end{frontmatter}

\section{Introduction}
To extend the HERA kinematics region at least by an order of magnitude in both $Q^{2}$ and $x_{g}$, the QCD Explorer collider was proposed \cite{engin03,sultansoy04,Schulte04}
by extrapolating earlier ideas of linac-ring type colliders
\cite{Csonka71,Sultanov89,Grosse89, ciftci01, sultansoy98} to a new kinematic range.
The QCD Explorer is a linac-ring type electron-proton collider making use
of a multi-GeV electron beam and the multi-TeV LHC proton or ion beam.
An obvious advantage of the QCD Explorer as compared to the ring-ring type e-p colliders (i.e. LHeC \cite{LHeC}) is the possibility to transform it into a
$\gamma p$ collider using the same infrastructure
\cite{ciftci95,sultansoy03,ciftci98,Aksakal05}.
This possibility is further explored in the present paper.

For the QCD Explorer the protons would be stored at the LHC design
energy of 7 TeV. The ions would have an equivalent energy, corresponding
to the same magentic bending field, and depending on their mass and charge
state.
The high energy electron beam could be either be the main-beam
equivalent of a single CLIC drive beam unit of CLIC
reaching about 75 GeV electron energy ("CLIC-1" \cite{asner03}) or
be produced by an ILC-type super conducting linac with an energy of 60 GeV.
In both cases, the final energy could be increased in stages by either
adding more drive beam units or further s.c.~cavities and klystrons,
respectively. The ultimate energy-frontier $\gamma$-nucleon collider
would employ a full 1.5-TeV CLIC linac colliding with the LHC
\cite{ciftci04}.

``CLIC-1" comprises a single drive-beam unit
which can accelerate the main beam to 75 GeV.
The bunch structure of "CLIC-1" does not well match the
nominal bunch structure of the LHC. The mismatch in bunch spacing
limits the achievable luminosity. Two remedial
approaches were proposed. In the first scheme, LHC operates
with a long superbunch, whoise length equals the length of the
CLIC bunch train. This schemes obviously requires
a change in the bunch structure of LHC proton beam \cite{frank04},
which could be realized in a number of ways, e.g., \cite{heikodamerau}.
Unfortunately, the superbunches are not compatible with
simultaneous running of the upgraded ATLAS and CMS detectors \cite{tapprogge}.
In the second approach the CLIC linac parameters are modified,
namely one considers a two times longer linac
(in one of three possible incarnations, called "CLIC-15a", "CLIC-15b",
and "CLIC-15c", which are detailed below) than "CLIC-1" with a
lower accelerating gradient (75 MV/m) and a two times
lower RF frequency (15 GHz), which will make it possible to change
the bunch charges, the bunch length
and the bunch separation time in the linac.

Namely, the bunch charge can be increased by a factor of two
for the reduced accelerating frequency.
The drive beam bunches arrive at a frequency of 15 GHz, so that
they can also drive 15 GHz accelerating structures.
For this purpose, one would use scaled version of the CLIC accelerating structures. The structure dimensions would all be doubled. The input power per structure would remain unchanged but the gradient is halfed. Due to the lower frequency the beam loading is reduced to a quarter of the original value. Taking into account the reduced gradient, this allows doubling the bunch charge. The distance between bunches needs to be doubled, leaving the beam current constant. The only drawback would be that the fill time for the main linac structures also doubles. Hence the number of bunches
is reduced from 220 to 92.
The corresponding beam parameters can be found in Table~\ref{table:1}. It should be noted that this mode of operation can create a problem with the beam loading compensation in the drive
beam accelerator.
In the current system two pulses are produced at the same time for a reason explained in the next paragraph.
If one wanted to avoid producing the second pulse,
the beam loading compensation scheme would need modification.
However, a simple means exists to avoid the problem and at the same time to increase the luminosity, as described below.

An improvement could be achieved by a modification of the delay loop of the drive beam generation complex. In the current scheme, this loop delays the trains by about 69.7 ns. This allow generating
trains of 69.7 ns length which are then combined in the subsequent system of combiner rings. In order to avoid too small rings a pair of trains is produced simultaneously. An increase of the delay loop
length to 139.4 ns would allow producing one pulse of twice the length instead. This would keep the ratio fill time to pulse length constant and allow to use 220 bunches per train at 15 GHz (``CLIC-15b"). This mode of operation has the advantage compared to the previous one that there will be no problem with the beam loading in the drive beam accelerator.

In the nominal CLIC (and hence also in ``CLIC-1")
one will not only produce a single drive beam pulse but rather a series of 22 pulses in order to power subsequent sections of the main linac.
Since the heat load induced by the RF system is
smaller at 15 GHz one could use more pulses per second than at 30 GHz. For the same Q-value this method would allow to increase the repetition rate by a factor two, assuming that we are limited by the
power transfer through the inner surface of the structure. The Q-value is expected to be larger at lower frequency scaling roughly as $\omega^{-1/2}$. This would allow to further increase the
repetition rate by about a factor of 1.4 (``CLIC-15c"). It should be noted that it may be possible to improve the structure design to increase this gain even further. Therefore we assume that three pulses are produced at each drive beam RF pulse with a spacing of 32$\times2\times$139.4 ns, where the factor 32 represents the nominal compression factor of the CLIC drive-beam complex.

Electron beam parameters for ``CLIC-1", ``CLIC-15a" and ``ILC" are summarized in
 Table~\ref{table:1}. ``CLIC-15b" has the same parameter set as "CLIC-15a" except for the number of bunches which is 220. ``CLIC-15c" has the same parameters as ``CLIC-15b", but a repetition frequency of 420 Hz instead of 150 Hz. Considering an interaction region of total length 200 cm, one proton bunch would collide with about 50 electron, or photon, bunches in the ``CLIC-1" option and with 25
bunches for ``CLIC-15a,b'' and ``c''. For the
``ILC" option,
each proton bunch would collide with a single photon bunch \cite{Aksakal05}.

In this study we consider a $\gamma$-nucleon collisions based on QCD
Explorer. In the $\gamma$-proton colliders, the high energy photons
could be produced by Compton backscattering of laser photons off a
high energy electron beam. To produce high energy photons, either
the ``ILC" or ``CLIC" options can be used. The Compton backscattered
photons collide with LHC's protons in the interaction region. In
Sections 2 and 3, we determine the electron beam and laser
parameters yielding an effective conversion. The achievable
luminosity for all cases is discussed in Section 4.

\section{Conversion region, Interaction region and beam parameters}
The conversion to high energy photon of a laser beam by colliding with an electron beam is determined by the Compton cross section. The total Compton cross section ($\sigma_{c}$) for polarized beams is
\cite{ginzburg83,ginzburg84}
\begin{equation}
\sigma_{c} =\sigma^{0} + 2 \lambda_{e} \lambda_{0} \sigma^{1},\\
\end{equation}
\begin{equation}
\sigma^{0} \;=\;\frac{2\pi\alpha^{2}}{xm_{e}^{2}}\;\left[\left(1-\frac{4}{x}-\frac{8}{x^{2}}\right)\ln\left(x+1\right)+\frac{1}{2}+\frac{8}{x}-\frac{1}{2\left(x+1\right)^{2}}\right],\\
\end{equation}
\begin{equation}
\sigma^{1}\;=\;\frac{2\pi\alpha^{2}}{xm_{e}^{2}}\;\left[\left(1+\frac{2}{x}\right)\ln\left(x+1\right)-\frac{5}{2}+\frac{1}{1+x}-\frac{1}{2\left(x+1\right)^{2}}\right],
\label{1}\\
\end{equation}
where $\alpha$ is the fine structure constant, $\lambda_{e}$ and $\lambda_{0}$ indicate the electron beam and laser beam helicities, respectively, and $x$ is a dimensionless parameter, which can be written as
\begin{equation}
x\;=\;\frac{4E_{b}\omega_{0}}{m^{2}}{\rm cos}^{2}\left(\frac{\alpha_{0}}{2}\right)\label{2}
\end{equation}
where $\alpha_{0}$ is the collision angle between laser and electron beams (in our calculation we will take it to be 0, corresponding to head-on collisions). The variables $E_{b}$ and $\omega_{0}$ denote the energy of the electron beam and laser photons. In the case of head-on collisions between the laser and electron beam, $x\simeq 15.3E_{b}\left[TeV\right] \omega_{0}\left[eV\right]$. The differential Compton cross section (for $\omega<\omega_{max}=E_b x/(x+1)$ reads\\
\begin{eqnarray}
 \frac{1}{\sigma_{c}}\frac{d\sigma_{c}}{d\omega}=f\left(\omega\right)=\frac{1}{E_{b}\sigma_{c}}
\frac{2\pi\alpha^{2}}{xm_{e}^{2}}
[\frac{1}{1-y}+1-y-4r\left(1-r\right)-\nonumber\\
\lambda_{e}\lambda_{0}rx\left(2r-1\right)\left(2-y\right)]\label{3}
\end{eqnarray}
where $y=\omega/E_{b}$ ($\omega$ is energy of backscattered
photons), and $r=y/[x(1-y)]$.

By varying the polarization of electron and laser beams, the
polarization of the high energy gamma beam can be tailored to fit
the needs of the gamma-proton/ion collision experiments. Controlling
the polarization is also important for sharpening the spectral peak
in the $\gamma p$ luminosity. Due to functional form of the Compton
scattering, the peak in the luminosity spectrum is significantly
enhanced by choosing the helicity of laser photons to be of opposite
sign to that of the electrons \cite{ginzburg84,Telnov90,Borden92}.

\subsection{Optimization of the laser parameters}
The maximum energy of the backscattered photons is
$\omega_{max}=\left[x/(x+1)\; E_{b}\right]$, depending on the parameter $x$ but the backscattered photons
can be lost for $x$ larger than 4.8 due to $e^{+}e^{-}$ pair creation in
collisions of the produced high-energy photons with the yet
un-scattered laser photons via the Breit-Wheeler process.
Thus, the optimum value is $x=4.8$, It translates into
the maximum photon energy $\omega_{max}=0.81 E_b$.
The angle of the backscattered photons with respect to the direction
of the incoming electron varies with photon energy as \cite{Borden92}
\begin{equation}
\theta_{\gamma}\left(\omega\right)\approx\frac{m_{e}}{E_{b}}\sqrt{\frac{E_{b}x}{\omega}-x+1}
\end{equation}
Neglecting multiple scattering, and assuming that the laser profile seen by each electron is the same, the conversion probability of generating high energy gamma photons per individual electron can be written as
\begin{equation}
p=1-e^{-q}
\end{equation}
If the laser intensity along the axis is uniform the exponent q is
\begin{equation}
q=\frac{A}{A_0}=\frac{\sigma_{c}A}{\omega_{0}\Sigma_{L}}=\frac{\sigma_{c}I\tau_{L}}{\omega_{0}}
\end{equation}
where A/$\omega_0$ denotes total number of laser photons, $\sigma_c$ is the total Compton cross section equal to 1.75 $10^{-25}\; {\rm cm}^{2}$ for $x=4.8$,
$I$ is the laser beam intensity and $\tau_{L}$ ($\sim\frac{2\sigma_{Lz}}{c}$) is the laser pulse length, $\Sigma_{L}=\frac{1}{2}\lambda Z_{R}$ the laser beam cross section at the focal point and $A$ is the laser pulse energy ($A=I\tau_{L}\Sigma_{L}$). The optimum conversion efficiency corresponds to q=1 which is reached for a laser pulse energy of $A=A_{0}=\omega_{o}\lambda Z_{R}/2\sigma_{c}$. In this case one has $p=0.65$.

The optimized laser-beam parameters for the ``ILC" and the
possible ``CLIC" options are listed in Table~\ref{table:2}.
It should be kept in mind that the transverse size of the laser beam
must be bigger than the electron beam size. The
laser beam size is defined by the final optical system of laser.
After the final optical element, the Rayleigh length is given by
\begin{equation}\label{}
    Z_{R}=\frac{4}{3}\lambda F_{N}
\end{equation}
where the $F_{N}$ value of the laser optics is defined as the ratio of
the focusing length of the last mirror to the incoming laser
beam diameter. The damage threshold of the mirror is taken to be
about 1 J.

\subsection{Beam parameters at conversion point and interaction region}
In Table~\ref{table:1} the beam parameters of the considered electron accelerators are given assuming a Gaussian beam distribution in all three spatial
dimensions. Since ``CLIC" and ``ILC" provide electron beams with different
energy, the laser parameters for ``CLIC" and ``ILC"
slightly differ at $x=4.8$. While an LHC proton bunch collides
only once with with a photon bunch produced from ``ILC",
50 photon-proton interaction points are considered over a
200-cm interaction region for the ``CLIC-1"$\times$LHC option.
For the other ``CLIC'' options we assume 25 collision points,
in view of the two times larger bunch spacing.

The LHC proton beam parameters are listed in the last column of Table ~\ref{table:1}. Electron and proton beam sizes along the s axis are given by
\begin{eqnarray}
\sigma_{j,i}(s)=\sigma_{j,i}^{*}\sqrt{1+\frac{(s-s_{j})^{2}}{(\beta_{j,i}^{*})^2}}
\label{eq:beamsize}.
\end{eqnarray}
where $j$ indicates the kind of beam
(e or p), $i$ ($=x,y$) the transverse coordinate,
$s_{j}$ the beam waist position,
and $\sigma_{j,i}^{*}$ ($=\sqrt{\varepsilon\beta}$) the transverse beam
size at the waist. Eq.~(\ref{eq:beamsize}) can be extended to the description of transverse laser beam sizes by changing $\beta$ with $Z_{R}$ and $\varepsilon$ with $\lambda/4\pi$. Here $Z_{R}$ is the Rayleigh length, $\lambda$ is the laser wavelength and $\sigma_{L,i}^{*}(=\sqrt{\frac{\lambda Z_{R}}{4\pi}})$ is the transverse laser beam size at the focal point. As stated before,
the distribution function the beam propagating in the
$z$ direction is assumed to be Gaussian in all three dimensions.

The distance between conversion point (CP) and interaction region (IR) is choosen as 75 cm, so at to be able to extract the spent electrons.
The transverse sizes of the electron beam are matched to
the proton beam sizes (11 $\mu m$) at the beginning
of the interaction region.

\section{Conversion efficiency}
\begin{figure}
\includegraphics[width=0.7\textwidth]{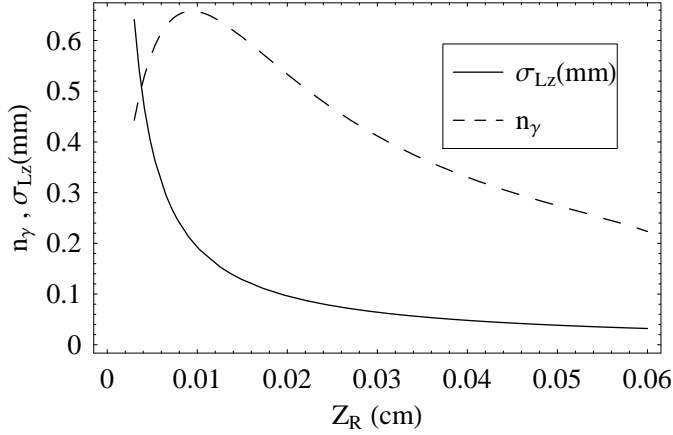}
\caption{Conversion efficiency and laser pulse length vs. $Z_{R}$ for ``CLIC-1"} \label{fig:conversion1}
\end{figure}

\begin{figure}
\includegraphics[width=0.7\textwidth]{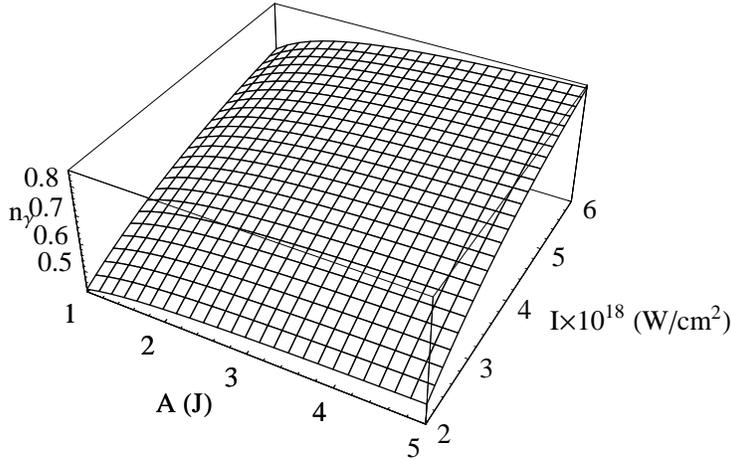}
\caption{Conversion efficiency vs laser pulse energy and intensity for ``CLIC-1"} \label{fig:conversion2}
\end{figure}

The conversion formula for the special case of head-on collision is
\begin{equation}
n_{\gamma}\equiv\frac{N_{\gamma}}{N_{e}}=1-\frac{1}{\sqrt{2\pi}\sigma_{ez}}
\int{exp\left(-\frac{z^{2}}{2\sigma_{ez}^{2}}-U\left(z\right)\right)dz}
\end{equation}
as given in Ref. \cite{NLC}. Where U(z) is
\begin{equation}
U\left(z\right)=\frac{4\sigma_{c}N_{L}}{\sqrt{2\pi}\lambda Z_{r}\sigma_{Lz}}\int \frac{exp\left(-\frac{2\left(s-\frac{z}{2}\right)^{2}}{\sigma_{Lz}^2}\right)}{1+\frac{s^{2}}{Z_{R}^{2}}}ds
\end{equation}
Here $N_{L}\left(=A/\omega_{0}\right)$ is the number of laser photons in the pulse, $\sigma_{ez}$ and $\sigma_{Lz}$ are the rms lengths of the electron bunch and of the laser pulse, respectively. Neglecting multiple scattering and assuming that the laser profile seen by each electron is the same, the optimum laser pulse length $\sigma_{Lz}$ and the conversion efficiency vary with $Z_{R}$ as seen in Figure~\ref{fig:conversion1}. Conversion efficiency is also obtained as a function of laser pulse energy and intensity as illustrated in Figure~\ref{fig:conversion2}.
The required laser pulse energy and intensity can also be inferred
from Figure~\ref{fig:conversion2}.

The electromagnetic field at the laser focus can give rise to
 multiphoton processes. The associated nonlinear effects are described by the parameter $\xi$. If $\xi^{2}\ll 1$, an electron interacts with one laser photon. Otherwise $\left(\xi^{2}\gg 1\right)$ multiphoton processes become dominant and the maximum photon energy decreases.
At the center of the conversion region, $\xi^{2}$ is given by
\begin{equation}
\xi^{2}=\frac{4r_{e}\lambda
A}{\left(2\pi\right)^{\frac{3}{2}}\sigma_{L,z}mc^{2}Z_{R}}\; ,
\end{equation}
where $r_{e}$ denotes the classical electron radius.
This equation imposes a lower limit on the Rayleigh length $(Z_{R})$.

\subsection{Extraction of spent electron beam}
There are several sources of electrons at the nucleon-photon IP.
Two of these are:

a) the initial electrons which are not scattered by laser photons at the CP;

b) the electrons which have lost part of their energy by Compton back scattering.

After conversion, the electrons with a wide energy spectrum can cross through a region with a transverse magnetic field B, where they are deflected in the orthogonal transverse direction. The deflection should be much larger than the proton beam size.

After conversion, the beam can be displaced at the IP. This
displacement $\tilde{y}=eBz^{2}/2E^{\prime}$ ($E^{\prime}$ being the
electron energy after collision). To provide
$\tilde{y}=10\sigma_{px}$ for un-scattered electrons of energy 75
GeV at a distance $z=75$~cm, one needs $B=0.98$~kG. To deflect
electrons which have suffered one Compton collision, the required
magnetic field is 0.18 kG on average. As mentioned earlier, in the
field of a high density laser, an electron may also undergo multiple
scattering. For such case, the electrons will have an even smaller
energy, and a correspondingly lower magnetic field would speep them
out. For the ``ILC" option, the deflection of the un-scattered
electrons produced at 75 cm from the main collision point requires a
magnetic field of 0.79 kG.

\section{Luminosity calculation}
Following Refs.~\cite{ciftci95,Aksakal05},
the equation describing the luminosity distribution is
\begin{equation}
\frac{dL_{\gamma
p}}{d\omega}=\frac{N_{\gamma}N_{p}f_{coll}f\left(\omega\right)}{
2\pi\left(\sigma_{e}^{2}+\sigma_{p}^{2}\right)}\exp\left[-\frac{z^{2}\theta_{\gamma}
\left(\omega\right)^2}{2\left(\sigma_{e}^{2}+\sigma_{p}^{2}\right)}\right]
\end{equation}
where $f\left(\omega\right)$ signifies the differential Compton cross section, $N_{\gamma}$ is the number of back scattered photons per pulse, $f_{coll}$ is collision frequency, $\theta\left(\omega\right)$ is angle of the backscattered photons, $\sigma_{e}$ and $\sigma_{p}$ are the transverse beam sizes of electrons and protons (or ions), respectively. Making a change of variables,
the $\gamma p$ luminosity distribution can be written in terms of the
invariant $\gamma p$ mass:
\begin{equation}
\frac{dL_{\gamma p}}{dW_{\gamma p}}=\frac{W_{\gamma
p}}{2E_{p}}\frac{N_{\gamma}N_{p}f_{coll}}{2\pi\left(
\sigma_{e}^{2}+\sigma_{p}^{2}\right)}f\left(\frac{W_{\gamma
p}^{2}}{4E{p}}\right)\exp
\left[-\frac{z^{2}\theta_{\gamma}\left(\frac{W_{\gamma
p}^{2}}{4E{p}}\right)^{2}}{2\left(
\sigma_{e}^{2}+\sigma_{p}^{2}\right)}\right]
\end{equation}
with $W_{\gamma p}=2\sqrt{E_{p}\omega}$ denoting invariant mass of
the $\gamma p$ system. The total luminosity of the  $\gamma p$
collisions is obtained by integration over the photon energy
$L_{\gamma p}=\int_{0}^{\omega_{max}} ( dL_{\gamma p}/ d\omega )\;
d\omega$ and summing over multiple interaction points. The total
$\gamma p$ luminosity for "CLIC-1"$\times$LHC at z=75 cm is
$1.55\times10^{29}\,cm^{-2}s^{-1}$ and other "CLIC" options are
$1.18\times 10^{29}$ for "CLIC-15a", $2.67\times10^{29}$ for
"CLIC-15b" and $7.5\times10^{29}$ for "CLIC-15c" and for "ILC" it is
$1.6\times10^{30}\,cm^{-2}s^{-1}$.

\begin{figure}
\includegraphics[width=0.7\textwidth]{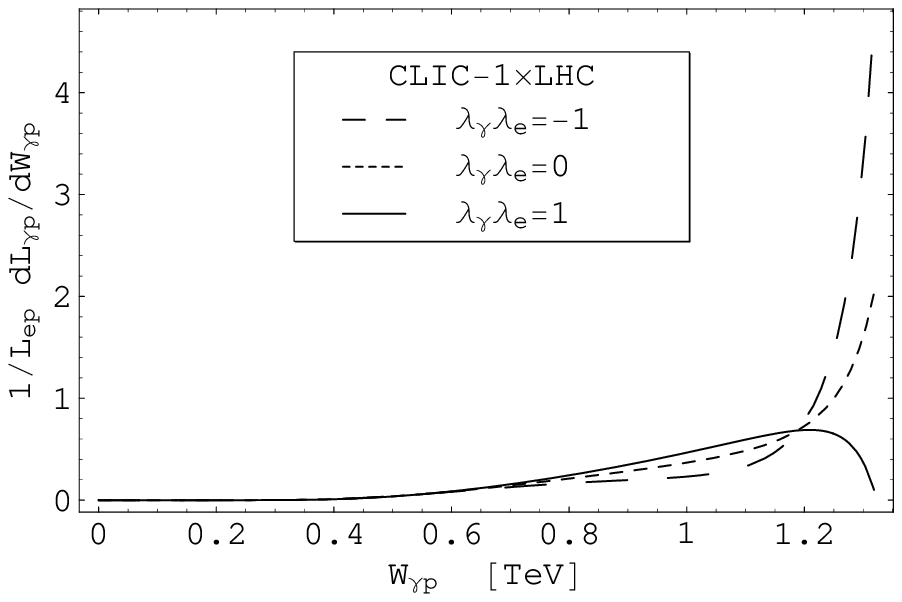}
\includegraphics[width=0.7\textwidth]{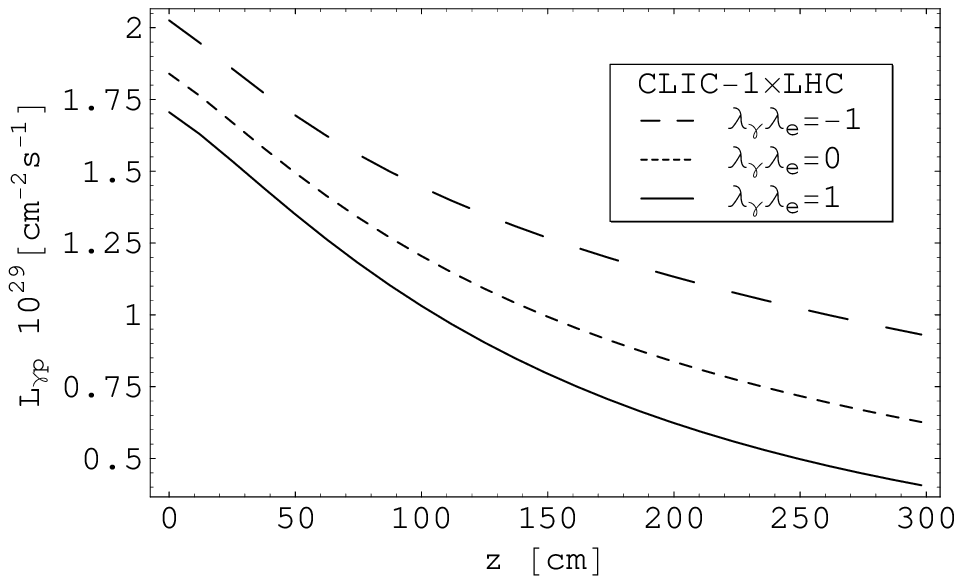}
\caption{ a)``CLIC-1"$\times$LHC Luminosity distribution for various
laser and electron helicities. b) ``CLIC-1"$\times$LHC total
luminosity vs.~$z$.} \label{fig:cliclumi1}
\end{figure}

\begin{figure}
\includegraphics[width=0.7\textwidth]{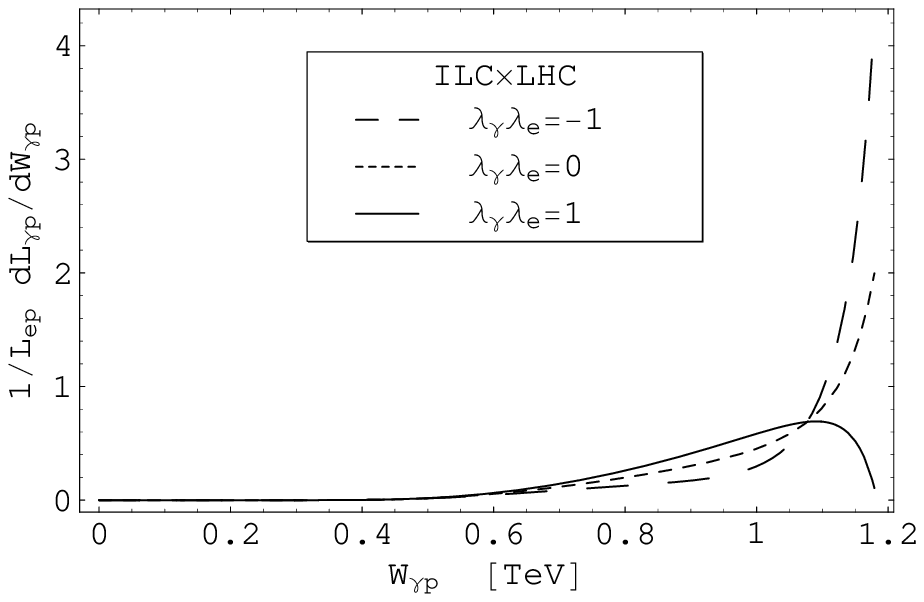}
\includegraphics[width=0.7\textwidth]{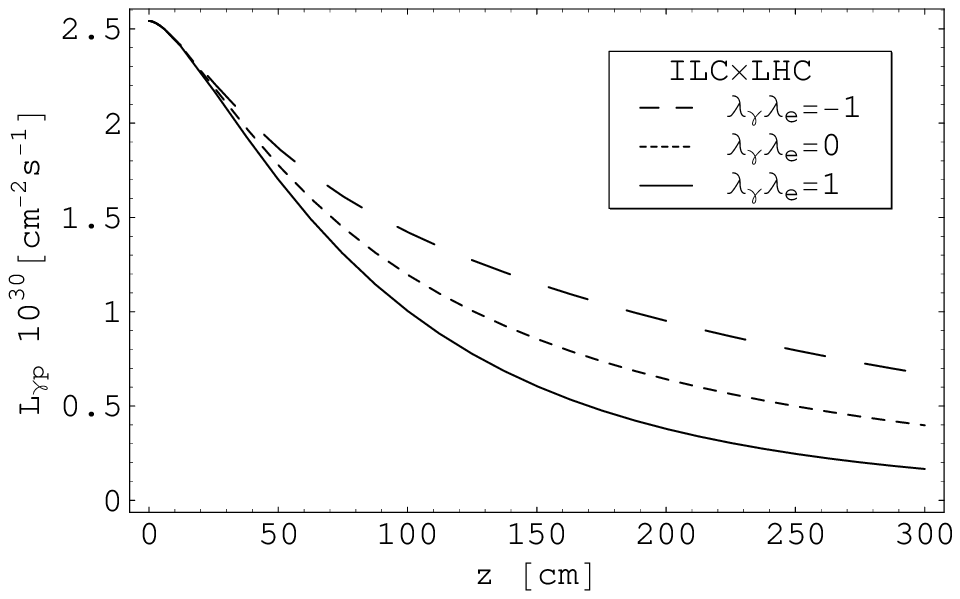}
\caption{ a) "ILC"$\times$LHC Luminosity distribution for various laser and electron helicities.
b) "ILC"$\times$LHC total luminosity vs. z}
\label{fig:lumi2}
\end{figure}

Figures~\ref{fig:cliclumi1} and ~\ref{fig:lumi2} show the luminosity of gamma-proton collider as a function of  the distance z and the invariant mass ($W_{\gamma p}$) for "CLIC-1"$\times$LHC and
"ILC"$\times$LHC respectively.

The luminosity for $\gamma p$ machine depends on the distance
between CP and IP (where z distance between $1^{st}$ IP and CP) and
also the laser and electron helicities. An increase of the distance
$z$ reduces the luminosity but also reduces the energy spread of the
photon beam.

\section{Physics goals}
A partial list of physics goals of $\gamma p$ colliders
based on the QCD Explorer concept includes \cite{sultansoy98,thera}:
\begin{itemize}
\item Total cross-section at TeV scale, which can be extrapolated from existing low energy data as $\sigma (\gamma p\rightarrow hadrons) \approx100\div 200\,\,\mu b$
\item Two-jet events, about $10^{4}$ events per working year with $p_{t}>100\,\,GeV$
\item Heavy quark pairs, $10^{7}\div 10^{8}\left(10^{6}\div10^{7},10^{2}\div10^{3}\right)$ events per operating
year for $c\bar{c}\left(b\bar{b},t\bar{t}\right)$ pair production
\item Hadronic structure of the photon
\item Single W production, $10^{4}\div10^{5}$ events per operating year
\item Single production of t-quark and fourth family quarks due to anomalous $\gamma-c-Q$ or $\gamma-u-Q\left(Q=t,\,u_{4}\right)$ and $\gamma-s-d_{4}$ or $\gamma-d-d_{4}$ interactions.
\end{itemize}

A preliminary list of physics goals of the
QCD Explorer based $\gamma A$ colliders comprises \cite{sultansoy98,thera}:
\begin {itemize}
\item total cross-section to clarify real mechanism of
very high energy $\gamma$-nucleus interactions;
\item investigation of a hadronic structure of the photon in nuclear medium;
\item according to the vector meson dominance (VMD) model,
the proposed machine will also be a $\rho$-nucleus collider;
\item formation of quark-gluon plasma at very high temperature but relatively low nuclear density;
\item the gluon distribution at extremly small $x_{g}$ in nuclear medium ($\gamma A\rightarrow QQ+X$);
\item investigation of both heavy quark and nuclear medium properties ($\gamma A\rightarrow J/\Psi(Y)+X,\,J/\Psi(Y)\rightarrow l^{+}l^{-}$);
\item existence of multi-quark cluster in nuclear medium and a few-nucleon correlation.
\end{itemize}

$\gamma A$ collider will give unique oportunity to investigate the small $x_{g}$ region in nuclear medium \cite{Frankfurt99}. Indeed, due to the advantage of the real $\gamma$ spectrum, heavy quarks will be produced via $\gamma g$
fusion at a characteristic $x$ parameter,
\begin{equation}
x_{g}\approx \frac{5\times m_{c(b)}^{2}}{0.8\times (Z/A)\times s_{ep}}\; ,
\end{equation}
which is approximately $\left(2-3 \right)\,10^{-5}$ for charmed and $\left(2-3 \right)\,10^{-4}$ for beauty hadrons. The number of $c\bar c$ and $b\bar b$ pairs which will be produced in $\gamma C$ collisions, can be estimated as $10^{6}-10^{7}$ and $10^{5}-10^{6}$ per working year, respectively. Therefore, one will be able to investigate the small $x_{g}$ region in detail. For this reason, a very forward detector in $\gamma$-beam direction will be useful for investigation of small $x_{g}$ region via detection of charmed and beauty hadrons.

\section{Conclusion}
Lepton-hadron collider with $\sqrt{s_{ep}}$ of order of 1 TeV is necessary both to clarify fundamental aspects of the QCD
part of the Standard Model and for adequate interpretation of experimental data from the LHC. Today, there are two
realistic proposals, namely, QCD Explorer and LHeC. Both QCD-E and LHeC will give opportunity to achieve sufficiently
high luminosity to explore crucial aspects of the strong interactions.
Even though values for luminosities of QCD-E with existing linac projects (ILC and CLIC) are lower than the value advertised for the LHeC one, the
luminosity of a QCD Explorer using a dedicated electron linac could
exceed that of the LHeC \cite{hande}.

In this paper, we have considered a $\gamma p$ collider based on QCD Explorer
with linac parameters taken from two existing linear
$e^{+}e^{-}$ collider designs, CLIC and ILC, with at most a few moderate
and straightforward modifications. Obviously, the luminosity of a $\gamma p$
collider would be higher for a QCD-Explorer with optimized linac parameters.
On the other hand, the competing LHeC proposal
requires the re-construction of an electron ring inside the LHC tunnel,
and must address the formidable problems of sharing the same tunnel with the
LHC proton ring including its rf sections and collimation regions,
of bypassing the huge detectors already installed around the
four LHC interaction points, and of proton-beam crab crossing.
It is also worth emphasizing that
the LHeC ring-ring collider cannot be transformed into a photon-nucleon
collider, while the QCD Explorer easily allows for this
extension, as described in this report. In addition,
the center of mass energy of the QCD-E based ep or $\gamma$p collider
can be increased simply by increasing the length of the electron linac,
while the energy of the LHeC is severely limited.

In our opinion one of the important features of a general
ep complex is the $\gamma A$ collider. Indeed, in the THERA
report \cite{thera} this type of collider was identified as the
most promising option for a TESLA$\times$HERA complex.

\section*{Acknowledgments}
The authors would like to thanks to Prof. Dr. Saleh Sultansoy and
Assoc. Prof. Dr. Gokhan Unel for useful discussions.

\newpage

\section*{Tables}
\begin{table}[!htb]
\caption{Beam Parameters of "CLIC-1", "CLIC-15a" (b/c), "ILC" and LHC ( Parameters in the paranthesis are used for "ILC"$\times$LHC collider)}
\label{table:1}
\begin{tabular}{lccccc}\hline
 Parameter                              & ``CLIC-1" & ``CLIC-15a (b/c)"   & ``ILC"   & LHC \\\hline
 Energy $E_{b}$ (GeV)                       & 75       & 75        & 60     & 7000 \\
 Bunch population $N_{b}$ $10^{10}$         & 0.256    & 0.512     & 2      & 17   \\
 RMS bunch length $\sigma_{z}$ ($\mu$m)     & 31       & 62     & 150   & 37.8 (75.5) mm\\
 Bunch spacing $t_{sep}$(ns)                & 0.267    & 0.534     & 300    & 5 (25) \\
 Number of bunches $n_{b}$                    & 220      & 92 (220 (b\&c))  & 2820   & 12 (2808)\\
 IP beta function  $\beta_{x,y}^{*}$(m)    & 26.8      & 26.8      & 14.1   & 0.25\\
 IP spot size $\sigma_{x,y}^{*}$($\mu$m)    & 11       & 11        & 11     & 11  \\
 CP beta function $\beta_{x,y}^{CP}$(cm)     & 2.1     & 2.1       & 4.0   & N/A\\
 CP spot size  $\sigma_{x,y}^{CP}$($\mu$m)  & 0.32     &0.32       & 0.58      & N/A \\
 Distance CP-IP $l_{CP-IP}$ (cm)            & 75        &75         & 75     & N/A\\
 RMS emittance $\gamma\varepsilon_{x,y}$($\mu mrad$)&0.7 &0.7       & 1      &3.75 \\
 Acc. Grad. (MV/m)                          &150       &75         & 35     & 3.75\\
 RF Freq.    (GHz)                          &30         &15         &15     &0.5 \\
 Repetition rate  $f_{rep}$ (Hz)             & 150      & 150 (420 (c))       & 5      &150\\
\hline
\end{tabular}
\end{table}

\begin{table}[!htb]
\caption {Laser Parameters for ``CLIC-1", ``CLIC-15a"(b/c), and ``ILC"}
\label{table:2}
\begin{tabular}{lccccc}\hline
 Parameter                              & "CLIC-1"       & "CLIC-15a"(b/c)    &"ILC" \\\hline
 Wavelength $\lambda$ ($\mu$m)             &0.296        &0.296           &0.240\\
 Pulse energy  A(J)                        &1            &1               &1  \\
 Rayleigh length  $Z_{R}$(mm)              &0.09         & 0.09           &0.1\\
 RMS spotsize at waist $\sigma_{L,i}^{*}$($\mu$m)&1.45   & 1.45           &2.17 \\
 RMS angular Divergence $\sigma_{L,i}^{'}$(mr)   &16.2   &16.2            &4    \\
 RMS pulse length  $\sigma_{Lz}$(mm)       &0.21         &0.21            &0.225\\
 Peak intensity I $10^{22}$ (Watt/$m^{2}$) &5.2          &5.2             &7.4 \\
 Nonlinear parameter $\xi^{2}$             &0.135         &0.135(0.115)   &0.054 \\\hline
\end{tabular}
\end{table}

\listoffigures

\end{document}